# The Security Assessment Domain: A Survey of Taxonomies and Ontologies

Ferrucio de Franco Rosa[1,2], Rodrigo Bonacin[1], Mario Jino[2]
[1]Renato Archer Information Technology Center (CTI), Campinas/SP – Brazil
[2]School of Electrical and Computer Engineering at University of Campinas (UNICAMP), Campinas, Brazil

*Abstract* – **The use of ontologies and taxonomies contributes by providing means to define concepts, minimize the ambiguity, improve the interoperability and manage knowledge of the security domain. Thus, this paper presents a literature survey on ontologies and taxonomies concerning the Security Assessment domain. We carried out it to uncover initiatives that aim at formalizing concepts from the "Information Security" and "Test and Assessment" fields of research. We applied a systematic review approach in seven scientific databases. 138 papers were identified and divided into categories according to their main contributions, namely: Ontology, Taxonomy and Survey. Based on their contents, we selected 47 papers on ontologies, 22 papers on taxonomies, and 11 papers on surveys. A taxonomy has been devised to be used in the evaluation of the papers. Summaries, tables, and a preliminary analysis of the selected works are presented. Our main contributions are: 1) an updated literature review, describing key characteristics, results, research issues, and application domains of the papers; and 2) the taxonomy for the evaluation process. We have also detected gaps in the Security Assessment literature that could be the subject of further studies in the field. This work is meant to be useful for security researchers who wish to adopt a formal approach in their methods and techniques.**

## I. INTRODUCTION

Theoretical knowledge and practical experience is available regarding the mechanisms and techniques aimed to improve information security created or acquired either in academia (state of the art) or in the software industry (state of the practice). However, this knowledge is dispersed, unstructured, non-systematic, or non-formalized.

Security defects, whose corrections are already known, continue to be introduced in new systems or exist in systems in operation, just waiting to be activated. On the other hand, it is not easy to detect and fix poorly defined security problems systematically. Activities to test the security of information systems often rely heavily on knowledge and experience of professionals involved in these activities [1]–[9].

Main research directions in Security Assessment and efforts aimed at formalizing the conceptual domain should be uncovered. A literature review is the tool to enables us to identify research issues, contributions, characteristics and objectives of these efforts.

Analysis of the selected works is as important as the literature review itself; the foundational concepts of the domain and their relationships must be uncovered. Domain knowledge is necessary to formalize research in a systematic way. The knowledge available in the literature supporting security assessment is not structured enough. Approaches based on a conceptual formalization, properly built into the information security context, can provide a better support for security assessment.

Hence, our main objective of this review is to search for initiatives aimed at formalizing concepts from the "Information Security" and "Test and Assessment" fields of research. To achieve this purpose, we performed a systematic review on the following databases IEEE Xplore, ACM Digital Library, Scielo, Proquest, ScienceDirect, SpringerLink, and Google Scholar. We also proposed and applied novel taxonomy to categorize the identified papers according to their main contribuitions. Selected papers are then individually analized and discussed in this article.

A preliminary evaluation of this study was presented in [10] focusing on ontologies. This article expands previous works by presenting ontologies and taxonomies, as well as a taxonomy to ease the evaluation of papers that address the Security Assessment area.

This paper is organized as follows: the second section presents the methodology used in the literature review; in the third section a taxonomy for analysis is proposed and the review is performed; in the fourth section we present related work; in the fifth section a discussion is conducted based on findings and comparisons; finally, we present the conclusions and proposals for future work.

## II. METHODOLOGY

Our literature review process is based, with adaptations, on the guidelines for performing systematic reviews proposed by Biolchini [11] and Kitchenham [12]. In the context of this study the literature review process is performed in six steps as follows: 1) Define secondary questions and the main research question; 2) Define keywords and search strings; 3) Define search databases; 4) Perform search and selection; 5) Extract relevant information from each paper; 6) Analyze the selected papers. The sixth step include the following tasks: 6.1) Define categories for classification of the papers; 6.2) Create summaries with key contributions of each paper; 6.3) Describe briefly the main characteristics of each paper; 6.4) Perform the evaluation of the quantitative and qualitative aspects. The information to be extracted from papers includes: research issues, main contributions, limitations, key characteristics, objectives and results.

## III. SURVEY

In this section, we present the literature review on "Information Systems Security Assessment". Specifically, we are looking for papers in the literature representing the state of the art in two search fronts. The first front aims to identify papers that systematize and formalize concepts on "Information Security", by means of taxonomies and ontologies. In the second front, we include "Test and Assessment" and the corresponding keywords.

### A. Defining secondary questions and the main research question

We start out with basic questions from which secondary questions are derived; these are summarized as the main research question. The basic questions are: (i) Which tools and techniques of the semantic web or from knowledge management can be used to formalize concepts and to define a common terminology in security assessment? (ii) How to use or map existing security ontologies? (iii) How to reuse knowledge in the security assessment process? (iv) How to reduce conceptual ambiguity in methods and techniques for security assessment? (v) How to reduce the dependence on expert knowledge in security evaluation? (vi) How to identify security properties covered by the evaluation? (vii) How to reduce the insecurity of systems, by means of feasible evaluation techniques?

From these basic questions, we derive secondary questions, to be answered from each search front, namely:

- *Search Front 1* – Conceptual formalization of Information Security. Search Front 1 aims to identify vocabularies, terminologies, taxonomies, ontologies and other forms of knowledge representation in Information Security, aimed to attain a consistent conceptual basis. Question 1 (Q1) – Which research efforts aim to support the systematization and formalization of knowledge in Information Security?
- *Search Front 2* – Assessment and Testing of Information Security. Search Front 2 aims to identify efforts on criteria and evaluation techniques based on a systematic use of conceptual formalization. Question 2 (Q2) – Which research efforts aim to assess or test Information Security properties in a systematic way?

From the secondary questions, we state the Main Research Question (MQ): Which approaches and techniques for knowledge formalization are applicable to the Security Information Assessment process?

### B. Defining keywords, search strings and search databases

For each question, we have defined keywords and search strings. English is used in the search. Table I presents the keywords and search strings.

TABLE I. KEYWORDS AND SEARCH STRINGS

| ***Question*** | ***Keywords*** | ***Search Strings*** |
|---|---|---|
| Q1 | Security; Privacy; Dependability; Ontology; Taxonomy. | ((Security OR Privacy OR Dependability) AND (Ontology OR Taxonomy)) |
| Q2, MQ | Security; Privacy; Dependability; Ontology; Taxonomy; Test; Testing; Assessment; Criterion; Criteria; Analysis; Audit; Evaluation. | ((Security OR Privacy OR Dependability) AND (Ontology OR Taxonomy) AND (Test OR Testing OR Assessment OR Criterion OR Criteria OR Analysis OR Audit OR Evaluation)) |

Seven databases were selected for this study: IEEE Xplore, ACM Digital Library, Scielo, Proquest, ScienceDirect – Elsevier, SpringerLink, and Google Scholar. The following parameters were considered: “title”, “abstract”, “entire document” and their combinations, if available in the search database.

### C. Performing Search and Selection

In the search and selection of papers the criteria used were: (i) Inclusion criteria: Recent works; Works with more citations; Works containing important concepts; Works related to the defined research questions. (ii) Exclusion criteria: Works with few citations, despite not being recent; Works not related to the defined research questions.

138 works of interest were identified. These papers were divided into categories according to their main contributions, namely: Ontology, Taxonomy and Survey. Based on their contents, 80 of them were selected: 47 on ontologies, 22 on taxonomies, and 11 on surveys (related works).

### D. Ontologies and Taxonomies

Ontologies can be used as a vocabulary, a dictionary or a roadmap of the Information Security domain. Furthermore, an ontology can be used to reason (provide inferences) about relationships between entities [13].

Ontologies are explicit specifications of conceptualizations [14]. Ontologies could be used to describe a particular context, providing meaning (semantics) to a vocabulary of words [15].

An ontology can be classified according to its degree of abstraction or generalization, as follows: (i) Top-level Ontology: it defines generic concepts that are independent of specific domains. More details can be found in [16]. (ii) Domain Ontology: it describes a particular domain, specifying concepts, properties and constraints. (iii) Task Ontology: it describes concepts of a task (or an action). (iv) Application Ontology: it describes concepts that consider both the domain context and the task context [15].

Albeit notations are available for ontology development, such as OWL or OntoUML, that is not the case for taxonomies. For instance, we can find papers presenting taxonomy as an extended vocabulary, a glossary, a list of concepts, concepts presented in a hierarchical way, etc. Hence, Guarino's classification [15] is used for analysis of both Ontologies (MC-OY) and Taxonomies (MC-TY), to ease the understanding of the analysis.

### E. Analysis of selected works

After the papers selection, an analysis was done addressing characterization, categorization, comparison, and inferences based on the data. Regarding the quantity of publications on ontologies and taxonomies per year, overall, there is a small quantity of taxonomies, particularly in recent papers. On the other hand, most of the recent papers address ontologies. Surveys are few as well as recent.

The selected works (Research Works – RW) were divided into categories according to their main contributions and research issues, and ranked in decreasing order by year of publication. As shown in Figure 1, the taxonomy for evaluation of the selected works is composed of categories, namely: Main Contribution (MC), Research Issue (RI), Application Domain (AD) and Characteristics (CR).

The MC of the selected research work is identified either by the author's description or by inference. A MC can be categorized as: Ontology (MC-OY), Taxonomy (MC-TY) or Survey (MC-SY).

The RI of the selected research work is identified either by the author's description or by inference. We can infer complements such as “need for...”, “difficulty of...”, “lack of...” and so on, preceding the RI's Description. RI can be: Formalizing knowledge (lack of domain's conceptualization, or conceptual ambiguity) (RI-FMK); Integrating and Interoperating Systems (RI-IOP); Reusing Knowledge (RI-RUS); Automating Processes (RI-APR); Increasing Coverage of Assessment (RI-CAS); Secure Sharing of Information (RI-SSI); Defining Security Standards (RI-DST); Detecting Intrusion (RI-DIN); Directioning Research (RI-DIR); Specifying Security Requirements (RI-REQ); Auditing Security (RI-AUD); Managing Knowledge (RI-KMN); Identifying Vulnerabilities (RI-IDV); Measuring Security (RI-MET); Suiting Processes or Methods (RI-PRO); Monitoring Security (RI-MNT); Establishing and Maintaining Security Policies (RI-POL); Designing Secure Systems (RI-DSS); Protecting Assets (RI-PTA); or Assessing, Verifying or Testing the Security (RI-AVS).

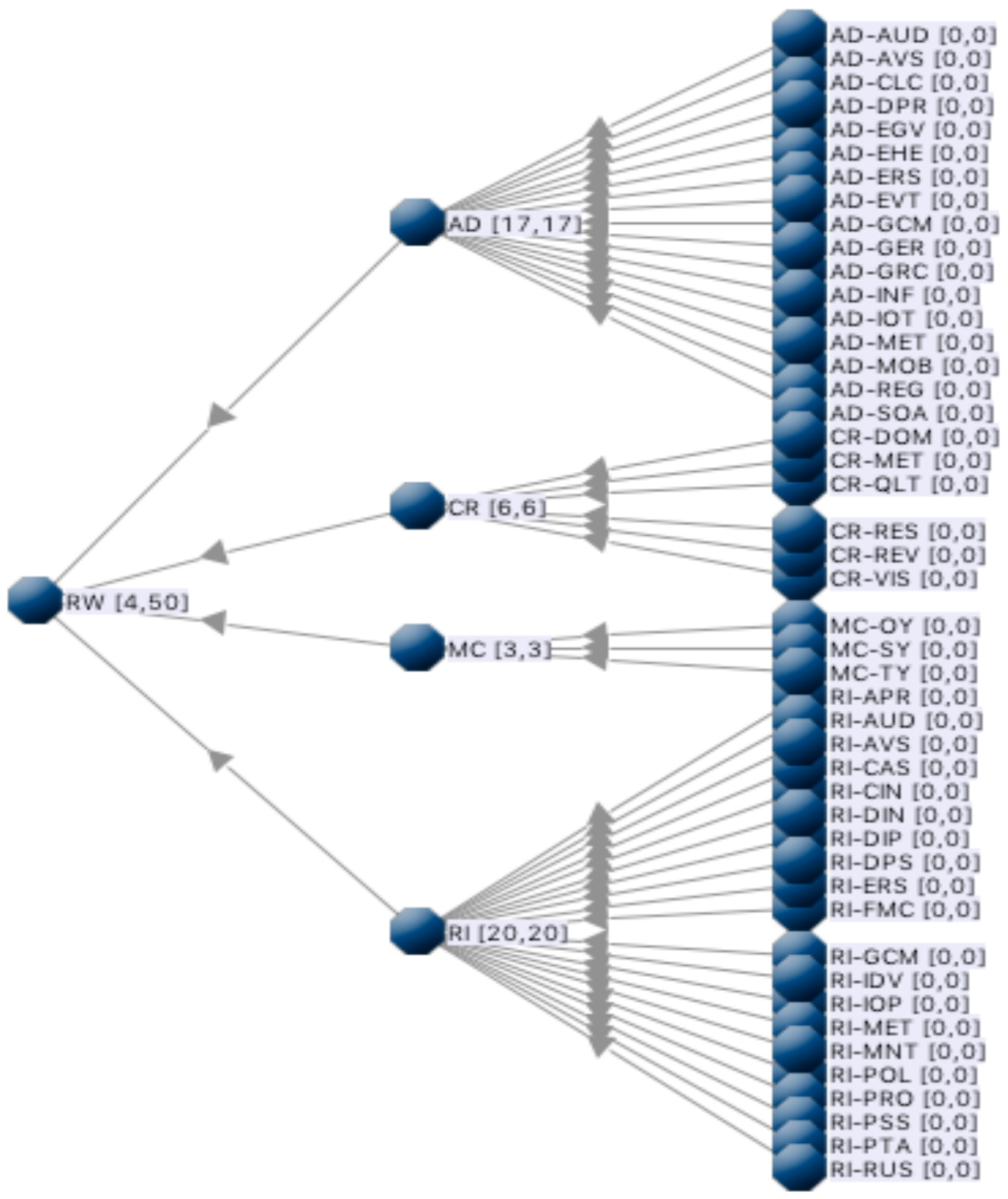

Fig. 1. Taxonomy for evaluation of the selected works.

The main AD of the selected research work is identified either by the author's description or by inference; particularly, whether the work addresses a specific scope of Information Security. AD can be: Security in General View (AD-GER); Requirements Specification (AD-REQ); Cloud Computing (AD-CLC); Knowledge Management (AD-KMN); Systems Assessment (AD-SAS); Infrastructure (AD-INF); E-Health (AD-EHE); Compliance Audit (AD-AUD); Metrics (AD-MET); E-Voting (AD-EVT); Service Oriented Architecture (SOA) (AD-SOA); E-government (AD-EGV); Risk Management (AD-RSK); Legal regulation (AD-REG); Detection and Prevention of Intrusions (AD-DPR); Embedded Systems and IoT (AD-IOT); or Mobile Applications (AD-MOB).

Key Characteristics (CR) of the works for which MC is survey (MC-SY) – Whether the work has the characteristics expected from a good quality review of the literature. CR can be: whether the work uses a systematic review method (CR-REV); whether the work briefly describes the systematic review method used, in order to enable repeatability (CR-MET); whether the work presents a brief summary of the main selected works (CR-SUM); whether the application domains of selected works are defined (CR-DOM); whether the work presents visualization tools for supporting the analysis. Visualization tools may be considered: tables, charts, taxonomies, etc. (CR-VIS); whether the work presents sufficient quantity of papers of good quality, enabling the reader to understand the domain (CR-QLT).

*F. Research Works on Ontologies (MC-OY)*

Most of the identified works aim to describe the domains of Software Security and Software Test, including their various sub-domains (e.g.: Risk Management, Security Policies; Incident Analysis; Attack Patterns, Performance Tests; Expert Systems Tests, etc.).

The 39 works that aim to describe domains and sub-domains are: [17]; [18]; [19]; [20]; [21], [22]; [23]; [24]; [25]; [26]; [27]; [28]; [29]; [30]; [31]; [32]; [33]; [34]; [35]; [36]; [37]; [38]; [39]; [40]; [41]; [42]; [43]; [44]; [45]; [46]; [47]; [48]; [49]; [50]; [51]; [50]; [13]; [52]; [53]. Generic and abstract proposals (Top-Level Ontologies) can be found in [54], [3], [55], [56], and [57]. Specific proposals (Task and Application Ontologies) can be found in [58], [59], and [60]. After analysis of the selected works, the following caracteristics can be highlighted:

Raskjn et al. (2002) [53] present concepts of the Information Security domain, and also explains how ontologies can be used to support the Information Security field, in order to provide a theoretical basis.

Viljanen (2005) [60] presents an ontology of Trust, in order to facilitate interoperability between systems. A common vocabulary is proposed to describe facts that should be considered in trust calculation.

Herzog et al. (2007) [13] present an information security ontology in OWL [61]. This work is aimed at modeling the main concepts of the domain. The authors describe content, manner of use, possibility of extension, technical implementation and tools to handle the ontology.

Fenz & Ekelhart (2009) [47] present an ontology of the Information Security domain, focused on Risk Management. The authors use the German IT Grundschutz Manual [62] and the NIST Handbook [63] as references. The ontology uses the concepts of threat, vulnerability and control to represent knowledge in the information security domain.

Evesti et al. (2011) [39] present an ontology to support the process of measuring Information Security.

Feledi & Fenz (2012) [28] present a formalization of information security knowledge, interpretable by machines, by means of a web portal (Web Protégé - [64]). According to the authors, we need to make explicit knowledge, so that it can be incorporated and used by both humans (human-readable format) and machines (machine-readable format).

Salini & Kanmani (2012) [3] present a top-level ontology of security requirements. Based on this ontology, we can design and develop requirements for electronic voting systems (e-voting). The main objective of this work is to propose security patterns to facilitate the process of identifying security requirements for e-voting systems. The authors present specific security properties for e-voting systems, namely: anonymity, disclosability, uniqueness, accuracy, transparency, and non-coercibility.

Gyrard et al. (2014) [25] present the STACK ontology (Security Toolbox: Attacks & Countermeasures) to aid developers in the design of secure applications. STACK defines security concepts such as attacks, countermeasures, security properties and their relationships. Countermeasures can be cryptographic concepts (encryption algorithm, key management, digital signature, and hash function), security tools, or security protocols. Kotenko et al. (2013) [24] present an ontology of security metrics, specifically built for the SIEM (Security Information and Event Management) domain.

Ramanauskaite et al. (2013) [23] presents a standard-based security ontology. After evaluating security ontologies, the authors conclude that the ontologies cover no more than one third of the standards. Thus, they propose a new ontology aimed to cover a larger number of standards. The authors mapped papers Herzog et al. (2007) [13] and Fenz, Pruckner, & Manutscheri (2009) [65] with the standards ISO 27001 [66], PCI DSS [67], ISSA 5173 [68] and NISTIR 7621 [69].

Salini & Kanmani (2013) [21] present an ontology of security requirements for web applications. This work aims at enabling the reuse of knowledge about security requirements in the development of different web applications.

Khairkar et al. (2013) [58] present an ontology to detect attacks on Web systems. The authors use semantic web concepts and ontologies to analyze security logs to identify potential security issues. This work aims to extract semantic relationships between attacks and intrusions in an Intrusion Detection System (IDS).

Kang & Liang (2013) [19] present a security ontology, for use in the software development process. The proposed ontology can be used for identifying security requirements, as a practical and theoretical basis.

Koinig et al. (2015) [17] present a security ontology for cloud computing and a brief literature review. The authors consider the regulatory requirements contained in standards such as HIPAA (Health Insurance Portability and Accountability Act) [70], SOX (Sarbanes Oxley Law) [71], and ISO/IEC 27001 [66].

### *G. Research Works on Taxonomies (MC-TY)*

16 works on taxonomies aim to describe domains and sub-domains: [72]; [73]; [74]; [75]; [76]; [77]; [78]; [79]; [80]; [81]; [82]; [83]; [84]; [85]; [86]; [87]. Both generic proposals (Top-Level Taxonomies) and specific proposals (Task and Application Taxonomies) can be found in [88], [89], [90], [91], [92], and [93]. After analysis of the selected works, the following caracteristics can be highlighted:

Wang & Wang (2003) [86] present a taxonomy of security risks. This work aims to conceptualize threats and risks to assess the security (wide view).

Avizienis et al. (2004) [85] present the main concepts of Dependability, including its attributes (e.g., confidentiality, integrity, authenticity, etc.) and the threats to those attributes. This work shows the relationships between the concepts of "dependability", "security", "survivability" and "trustworthiness", which are usually described as synonymous.

Clark et al. (2005) [84] present a taxonomy to categorize security attacks. The proposed taxonomy can provide a mechanism to infer on the probable incomplete aspects of a model. Roosta et al. (2006) [83] present a taxonomy of attacks to sensor networks and proposes solutions (countermeasures).

Savola (2007b) [88] presents a top-level taxonomy of information security metrics. A survey of the main approaches of security metrics is presented. The academic, governmental and industrial perspectives are considered.

Hu et al. (2008) [81] present a taxonomy of information security. According to the authors, there are two research communities working separately, namely: "Dependability" and "Security". The authors explain that the community of "Dependability" is more concerned about non-malicious failures and the community of "Security" is oriented to identify malicious failures and attacks. This work aims to integrate the two views into a single taxonomy. First, the authors describe the concept of "Dependability", subdivided into attributes, threats and faults. After integration with the attributes of the "Security" concept, the authors propose a conceptual framework. The framework is a set of the following attributes: Availability, Reliability, Integrity, Safety, Maintainability, Confidentiality, Authenticity, and Non-repudiation. The authors also present a classification of faults made by humans (HMF - Human-made faults), categorized into FUA (Faults with Unauthorized Access) and NFUA (Non-FUA - other faults). Subsequently, for each property, the structure of the FUA fault type is detailed.

Friedman & Hoffman (2008) [80] present a taxonomy of threats and defenses. The paper presents the threats to mobile devices and their data, as well as the available defenses. The taxonomy divides the threats to mobile devices into seven categories, namely: malware; phishing and social engineering; direct hacker attacks; interception and spoofing of communication; loss and theft of the device; malicious use of internal actions; policy violations. Countermeasures are divided into: firewall; antivirus and zero day antimalware software; IPS; VPN; data encryption technologies for data leak prevention. The authors emphasize that the devices move out of the protected perimeter, connected to unknown environments, and they can also be lost or stolen.

Sahoo et al. (2010) [78] present a taxonomy of virtual machines vulnerabilities, including concepts and problems associated with these vulnerabilities. We can see that there are common characteristics with the current term "Cloud Computing".

Evesti & Pantsar-Syväniemi (2010) [77] present a security taxonomy, in the context of mobile application environments. Current systems are intended for use in mobile or embedded devices and the environments changes during the execution of the application. Any changes in the operating environment (IT infrastructure) could cause security threats. For mobile applications, the operating environment is sometimes unknown or unstable. According to the authors, the software has to be constantly aware of its security level.

Babar et al. (2010) [76] provide an overview, analysis and taxonomy of security and privacy challenges in the Internet of Things – IoT. The authors propose a security model for IoT. This paper presents: key properties for IoT; the specific challenges of the IoT domain; high-level security requirements for IoT; a definition of resilience to attacks.

Paintsil & Fritsch (2011) [74] present a taxonomy of security risks. The taxonomy is based on a literature review of articles on IDMS (IDentity Management System). The taxonomy addresses the aspect of "authentication"; more specifically, how authentication tokens affect the privacy and security of IDMS.

Mundie & Mcintire (2013) [73] proposes a Malware Analysis taxonomy. This paper presents the results of a project that aims to develop a common vocabulary of Malware Analysis (Malware Analysis Lexicon - MAL). According to the authors, there is not a controlled vocabulary for malware analysis because there are many local dialects and a lack of standardization of the concepts. This work can be used as a source for building security ontologies due to a glossary of terms.

Wangen & Snekkenes (2014) [72] present a taxonomy of research issues in the Risk Management field. This paper aims to present the challenges for risk management of information security.

## IV. RELATED WORK

In this section, we analyse works that present a survey as the main contribution (MC-SY). After evaluating and classifying the works, we highlight the works by Blanco [94], [95], Mellado [96] and Souag [97], as they all present characteristics expected of a good quality literature review (CR-REV, CR-MET, CR-SUM, CR-DOM, CR-VIS, CR-QLT).

Blanco et al. (2008) [94] and Mellado et al. (2010) [96] were excluded from the comparison for not being recent (up to 2011). Blanco et al. (2011) [95] and Souag et al. (2012) [97] present useful comparisons but need updating.

Blanco et al. (2011) [95] present a literature review and proposes a method for integrating ontologies, through qualitative analysis of more mature proposals. Souag et al. (2012) [97] present an analysis of existing security ontologies and their use in defining requirements. The work is part of a project that aims to improve the definition of security requirements using ontologies. This study addresses the question: which security ontology is suitable for my needs? Ontologies are classified into eight families, by extending previous works. The authors analyzed how each ontology covers each aspect of security (e.g., Goals, assets, vulnerabilities, threats, and countermeasures). The literature review was adapted from Barnes (2005) [98] and Rainer & Miller (2005) [99]. The authors have also used Blanco et al. (2008) [94], Elahi (2009) [100] and Nguyen (2011) [101] as references.

Blanco et al. (2011) [95] and Souag et al. (2012) [97] emphasize the importance of previous literature reviews and point to the need of updates.

Table II presents a brief comparison among the most recent related works. All retrieved papers that present a survey as the main contribution (MC-SY) are shown in Appendix III.

TABLE II. RELATED WORK

| *Ref.* | *Authors* | *Particularities and differences* | *AD* | *MC* |
|---|---|---|---|---|
| [95] | (Blanco et al., 2011) | It focuses on managing security knowledge and on integrating security ontologies; A specific assessment of security ontologies (OntoMetrics) was performed; It does not consider taxonomies in the selection of the research works. | AD-KMN | MC-SY |
| [97] | (Souag; Salinesi; Comyn-Wattiau, 2012) | It focuses on specifying security requirements; It presents a classification into families of security ontologies; It considers taxonomies in the selection of research works. | AD-REQ | MC-SY |
| Our | Our | It focuses on the security assessment domain; It contains research works from 2012; It considers taxonomies in the selection of the research works; It presents a taxonomy for evaluating the research works. | AD-SAS | MC-SY; MC-TY |

## V. DISCUSSION

Most of the works address the conceptual formalization issue (RI-FMK). RI-FMK was identified in 100% of the papers on ontologies up to 2008. After 2008, we can see other applications for ontologies.

Concerning the ontologies (MC-OY), we can identify that among recent works (2011-2015) most are still RI-FMK. However, we can identify works that address other research issues, namely: RI-REQ [22], [3], [33], [35]; RI-AUD [23], [27], [29], [30]; RI-KMN [18], [28], [40], [46].

We can see gaps in the literature. There is a lack of works that address the following RIs: RI-RUS, RI-APR, RI-CAS RI-SSI, RI-DST, RI-IDV, RI-MET, RI-PTA, RI-AVS. This finding indicates that these RIs can be addressed in future papers, increasing the likelihood of original contributions.

Concerning AD, there is a good distribution among the domains in recent works. Most of the least recent works (up to 2009) are: AD-GER and AD-KMN. On the other hand, AD-REG, AD-DPR, AD-IOT and AD-MOB are not found in recent works (from 2011 on).

Regarding taxonomies (MC-TY), most of the RIs is RI-FMK, with the exception of [77] (RI-ERS), [80] (RI-MNT) and [89] (RI-MET).

There were not works addressing the following ADs: AD-EHE, AD-AUD, AD-EVT, AD-SOA, AD-EGV, and AD-REG. This finding indicates that studies applied in these ADs can be addressed in future works, increasing the likelihood of original contributions.

Recent works address a wide variety of application domains (AD). This finding could indicate that attempts to describe newer and smaller research fields may be a feasible alternative [13], [20], [47]. In Appendices I, II and III we present comparative tables of the papers related to ontologies and taxonomies.

## VI. CONCLUSION

We have presented a literature review of ontologies and taxonomies of the Security Assessment domain to search for initiatives aimed at formalizing concepts from the "Information Security" and "Test and Assessment" fields.

We identified gaps in the Security Assessment literature. There is a lack of works that address the following research issues: Reusing Knowledge; Automating Processes; Increasing Coverage of Assessment; Secure Sharing of Information; Defining Security Standards; Identifying Vulnerabilities; Measuring Security; Protecting Assets; Assessing, Verifying or Testing the Security. This finding indicates that these RIs can be addressed in future papers, increasing the likelihood of original contributions.

We applied a systematic approach in the literature review process by selecting search databases, choosing keywords, defining search strings, and specifying inclusion and exclusion criteria. Summaries, tables, and a preliminary analysis of the selected research papers are presented. The 138 identified papers were divided into categories according to their main contributions, namely: Ontology, Taxonomy and Survey. Based on their contents, 80 of them were selected including: 47 on ontologies, 22 on taxonomies, and 11 on surveys.

The literature review shows key characteristics, research results, research issues, application fields of the works, and also a categorization (taxonomy). The taxonomy for analyzing the research works may be extended, by including new main contributions (MC), characteristics (CR), application domains (AD), or research issues (RI).

This work is meant to be useful for security researchers who wish to formalize knowledge in their methods and techniques. We are currently working on expanding this survey and the taxonomy, to include other main contributions, such as "Methods" (MC-MT), "Approaches" (MC-AP), "Systems" (MC-ST), amng other contributions to be identified and categorized.

*APPENDIX I – Main Contribution – Ontology (MC-OY)*

| | RI | | | | | | | | | | | | | | | | | | | |
|---|---|---|---|---|---|---|---|---|---|---|---|---|---|---|---|---|---|---|---|---|
| Ref. | FMK | IOP | RUS | APR | CAS | SSI | DST | DIN | DIR | REQ | AUD | KMN | IDV | MET | PRO | MNT | POL | DSS | PTA | AVS |
| [54] | X | | | | | | | | | | | | | | | | | | | |
| [17] | X | | | | | | | | | | | | | | | | | | | |
| [18] | | | | | | | | | | | | X | | | | | | | | |
| [19] | X | | | | | | | | | | | | | | | | | | | |
| [20] | X | | | | | | | | | | | | | | | | | | | |
| [58] | | | | | | | | X | | | | | | | | | | | | |
| [21] | X | | | | | | | | | | | | | | | | | | | |
| [22] | | | | | | | | | | X | | | | | | | | | | |
| [23] | | | | | | | | | | | X | | | | | | | | | |
| [24] | X | | | | | | | | | | | | | | | | | | | |
| [25] | X | | | | | | | | | | | | | | | | | | | |
| [3] | | | | | | | | | | X | | | | | | | | | | |
| [26] | | | | | | | | | | | | | | | | | | X | | |
| [27] | | | | | | | | | | | X | | | | | | | | | |
| [28] | | | | | | | | | | | | X | | | | | | | | |
| [55] | | | | | | | | | | | | | | | | | X | | | |
| [29] | | | | | | | | | | | X | | | | | | | | | |
| [30] | | | | | | | | | | | X | | | | | | | | | |
| [31] | X | | | | | | | | | | | | | | | | | | | |
| [32] | | | | | | | | | | | | | | | X | | | | | |
| [33] | | | | | | | | | | X | | | | | | | | | | |
| [34] | X | | | | | | | | | | | | | | | | | | | |
| [35] | | | | | | | | | | X | | | | | | | | | | |
| [36] | X | | | | | | | | | | | | | | | | | | | |
| [37] | | | | | | | | | | | | | | | | X | | | | |
| [38] | | X | | | | | | | | | | | | | | | | | | |
| [39] | X | | | | | | | | | | | | | | | | | | | |
| [40] | | | | | | | | | | | | X | | | | | | | | |
| [41] | | | | | | | | | | | | | | | | | X | | | |
| [42] | X | | | | | | | | | | | | | | | | | | | |
| [43] | | | | | | | | X | | | | | | | | | | | | |
| [59] | X | | | | | | | | | | | | | | | | | | | |
| [44] | X | | | | | | | | | | | | | | | | | | | |
| [45] | X | | | | | | | | | | | | | | | | | | | |
| [46] | | | | | | | | | | | | X | | | | | | | | |
| [47] | X | | | | | | | | | | | | | | | | | | | |
| [48] | | | | | | | | | | | | | | | X | | | | | |
| [49] | | | | | | | | | | | | X | | | | | | | | |
| [50] | X | | | | | | | | | | | | | | | | | | | |
| [102] | X | | | | | | | | | | | | | | | | | | | |
| [51] | X | | | | | | | | | | | | | | | | | | | |
| [13] | X | | | | | | | | | | | | | | | | | | | |
| [52] | X | | | | | | | | | | | | | | | | | | | |
| [56] | X | | | | | | | | | | | | | | | | | | | |
| [60] | X | | | | | | | | | | | | | | | | | | | |
| [57] | X | | | | | | | | | | | | | | | | | | | |
| [53] | X | | | | | | | | | | | | | | | | | | | |

| | AD | | | | | | | | | | | | | | | | |
|---|---|---|---|---|---|---|---|---|---|---|---|---|---|---|---|---|---|
| Ref. | GER | REQ | CLC | KMN | SAS | INF | EHE | AUD | MET | EVT | SOA | EGV | RSK | REG | DPR | IOT | MOB |
| [54] | | X | | | | | | | | | | | | | | | |
| [17] | | | X | | | | | | | | | | | | | | |
| [18] | | | | | X | | | | | | | | | | | | |
| [19] | | | | X | | | | | | | | | | | | | |
| [20] | | | | | X | | | | | | | | | | | | |
| [58] | | | | | | X | | | | | | | | | | | |
| [21] | | X | | | | | | | | | | | | | | | |
| [22] | | | | | | | X | | | | | | | | | | |
| [23] | | | | | | | | X | | | | | | | | | |
| [24] | | | | | | | | | X | | | | | | | | |
| [25] | | | | X | | | | | | | | | | | | | |
| [3] | | | | | | | | | | X | | | | | | | |
| [26] | | X | | | | | | | | | | | | | | | |
| [27] | | | | | | | | | | | X | | | | | | |
| [28] | | | | X | | | | | | | | | | | | | |
| [55] | | | | | | | | | | | | X | | | | | |
| [29] | | | | | | | | | | | | | X | | | | |
| [30] | X | | | | | | | | | | | | | | | | |
| [31] | | | X | | | | | | | | | | | | | | |
| [32] | | | | | | | | | | | X | | | | | | |
| [33] | | | | | | | | | | | X | | | | | | |
| [34] | | | | X | | | | | | | | | | | | | |
| [35] | | | | | X | | | | | | | | | | | | |
| [36] | | | | | | X | | | | | | | | | | | |
| [37] | | | | | | X | | | | | | | | | | | |
| [38] | | | | | | | X | | | | | | | | | | |
| [39] | | | | X | | | | | | | | | | | | | |
| [40] | | | | X | | | | | | | | | | | | | |
| [41] | | X | | | | | | | | | | | | | | | |
| [42] | | | | | | X | | | | | | | | | | | |
| [43] | | | | | | X | | | | | | | | | | | |
| [59] | | | | | | | | | | | | | X | | | | |
| [44] | | | X | | | | | | | | | | | | | | |
| [45] | | | | | | | | | X | | | | | | | | |
| [46] | | | | X | | | | | | | | | | | | | |
| [47] | | | | | | | | | | | | | X | | | | |
| [48] | | | | X | | | | | | | | | | | | | |
| [49] | | | | X | | | | | | | | | | | | | |
| [50] | X | | | | | | | | | | | | | | | | |
| [102] | X | | | | | | | | | | | | | | | | |
| [51] | | | | X | | | | | | | | | | | | | |
| [13] | X | | | | | | | | | | | | | | | | |
| [52] | | | | X | | | | | | | | | | | | | |
| [56] | | | | X | | | | | | | | | | | | | |
| [60] | X | | | | | | | | | | | | | | | | |
| [57] | | | | | | | | | | | | | | X | | | |
| [53] | X | | | | | | | | | | | | | | | | |

*APPENDIX II – Main Contribution – Taxonomy (MC-TY)*

| | AD | | | | | | | | | | | | | | | | |
|---|---|---|---|---|---|---|---|---|---|---|---|---|---|---|---|---|---|
| **Ref.** | **GER** | **REQ** | **CLC** | **KMN** | **SAS** | **INF** | **EHE** | **AUD** | **MET** | **EVT** | **SOA** | **EGV** | **RSK** | **REG** | **DPR** | **IOT** | **MOB** |
| [72] | | | | | | | | | | | | | **X** | | | | |
| [73] | **X** | | | | | | | | | | | | | | | | |
| [74] | | | | | | | | | | | | | **X** | | | | |
| [75] | | | | | | | | | | | | | | | **X** | | |
| [76] | | | | | | | | | | | | | | | | **X** | |
| [77] | | | | | | | | | | | | | | | | | **X** |
| [78] | | | **X** | | | | | | | | | | | | | | |
| [79] | | | | | | | | | | | | | | | **X** | | |
| [80] | | | | | | | | | | | | | | | | | **X** |
| [81] | | | | **X** | | | | | | | | | | | | | |
| [89] | | | | | | | | | **X** | | | | | | | | |
| [88] | | | | | | | | | **X** | | | | | | | | |
| [82] | | | | | | **X** | | | | | | | | | | | |
| [83] | | | | | | | | | | | | | | | **X** | | |
| [84] | | | | | | | | | | | | | | | **X** | | |
| [85] | **X** | | | | | | | | | | | | | | | | |
| [90] | **X** | | | | | | | | | | | | | | | | |
| [86] | | | | | **X** | | | | | | | | | | | | |
| [91] | | **X** | | | | | | | | | | | | | | | |
| [92] | | | | | | **X** | | | | | | | | | | | |
| [93] | | | | | **X** | | | | | | | | | | | | |
| [87] | | | | **X** | | | | | | | | | | | | | |

*APPENDIX III – Main Contribution – Survey (MC-SY)*

| | MC | | | CR | | | | | | AD | | | | | | | | | | | | | | | | |
|---|---|---|---|---|---|---|---|---|---|---|---|---|---|---|---|---|---|---|---|---|---|---|---|---|---|---|
| **Ref.** | **OY** | **TY** | **SY** | **REV** | **MET** | **SUM** | **DOM** | **VIS** | **QLT** | **GER** | **REQ** | **CLC** | **KMN** | **SAS** | **INF** | **EHE** | **AUD** | **MET** | **EVT** | **SOA** | **EGV** | **RSK** | **REG** | **DPR** | **IOT** | **MOB** |
| [103] | | | **X** | | | **X** | | | | | | | **X** | | | | | | | | | | | | | |
| [104] | | | **X** | | | **X** | **X** | **X** | **X** | | | **X** | | | | | | | | | | | | | | |
| [105] | | | **X** | | | **X** | **X** | | **X** | **X** | | | | | | | | | | | | | | | | |
| [106] | | | **X** | | | **X** | **X** | **X** | **X** | | | | **X** | | | | | | | | | | | | | |
| [97] | | | **X** | **X** | **X** | **X** | **X** | **X** | **X** | | **X** | | | | | | | | | | | | | | | |
| [95] | | | **X** | **X** | **X** | **X** | **X** | **X** | **X** | | | | **X** | | | | | | | | | | | | | |
| [96] | | | **X** | **X** | **X** | **X** | **X** | **X** | **X** | | **X** | | | | | | | | | | | | | | | |
| [107] | | | **X** | | | **X** | **X** | **X** | **X** | **X** | | | | | | | | | | | | | | | | |
| [108] | | | **X** | | | **X** | **X** | **X** | **X** | | | | **X** | | | | | | | | | | | | | |
| [94] | | | **X** | **X** | **X** | **X** | **X** | **X** | **X** | **X** | | | | | | | | | | | | | | | | |
| [109] | | | **X** | | | **X** | **X** | **X** | **X** | **X** | | | | | | | | | | | | | | | | |
| ***Our*** | | ***X*** | ***X*** | ***X*** | ***X*** | ***X*** | ***X*** | ***X*** | ***X*** | | | | | ***X*** | | | | | | | | | | | | |